\documentclass[5p, twocolumn, sort&compress]{elsarticle}
\usepackage[utf8]{inputenc}
\usepackage{microtype}
\usepackage{xspace}
\usepackage{mathrsfs}
\usepackage{amsfonts, mathtools, mathrsfs, cuted}
\usepackage{bm}
\usepackage{xcolor}
\usepackage[colorlinks=true, linkcolor = blue, hyperfootnotes=false, citecolor = blue]{hyperref}
\newcommand{\R}{\mathbb{R}}
\newcommand{\norm}[1]{\left\lVert #1 \right\rVert}

\newcommand{\scalar}[1]{\left\langle{#1}\right\rangle}
\newcommand{\dd}{\mathrm{d}}
\renewcommand{\t}{\mathbf}
\newcommand{\gt}{\bm}
\renewcommand\[{\begin{equation}}
\renewcommand\]{\end{equation}}

\newcommand{\highlight}[1]{#1}

\begin{document}
\begin{frontmatter}
\title{Strain compatibility and gradient elasticity in morphing origami metamaterials}
\author[1]{Hussein Nassar}\ead{nassarh@missouri.edu}
\author[2]{Arthur Lebée}\ead{arthur.lebee@enpc.fr}
\author[1]{Emily Werner}

\address[1]{Department of Mechanical and Aerospace Engineering, University of Missouri, Columbia, MO 65211, USA}
\address[2]{Laboratoire Navier, École des Ponts, Université Gustave Eiffel, CNRS, France}

\begin{abstract}
The principles of origami design have proven useful in a number of technological applications. Origami tessellations in particular constitute a class of morphing metamaterials with unusual geometric and elastic properties. Although inextensible in principle, fine creases allow origami metamaterials to effectively deform non-isometrically. Determining the strains that are compatible with coarse-grained origami kinematics as well as the corresponding elasticity functionals is paramount to understanding and controlling the morphing paths of origami metamaterials. Here, within a unified theory, we solve this problem for a wide array of well-known origami tessellations including the Miura-ori as well as its more formidable oblique, non-developable and non-flat-foldable variants. We find that these patterns exhibit two universal properties. On one hand, they all admit equal but opposite in-plane and out-of-plane Poisson's ratios. On the other hand, their bending energy detaches from their in-plane strain and depends instead on the strain gradient. The results are illustrated over a case study of the self-equilibrium geometry of origami pillars.
\end{abstract}

\end{frontmatter}
\section{Introduction}
Origami has long surpassed its artistic vocation. Its early uses in architectural geometry date back to the 1920s \cite{wingler1978bauhaus, Lebee2015}. By now, origami techniques have been applied to the design of deployable structures in astronautics, robotics and biomedical engineering \cite{Kuribayashi2006, Schenk2011a, Zirbel2013, Callens2017}. Origami tessellations in particular make up versatile morphing 2D metamaterials that can access various low-energy high-curvature configurations \cite{Seffen2012,norman2009phd}. 
Although inextensible in principle, the fine crease patterns of origami metamaterials enable them to effectively deform non-isometrically. In this letter, we determine the strains that are compatible with coarse-grained origami kinematics. Strain compatibility permits to fully characterize the 3D-embedded geometry when suitable boundary conditions are prescribed. In most cases however, the geometry further solves an elastostatic equilibrium for which we calculate the appropriate elasticity functional.

\begin{figure}[ht!]
    \centering
    \includegraphics[width=\linewidth]{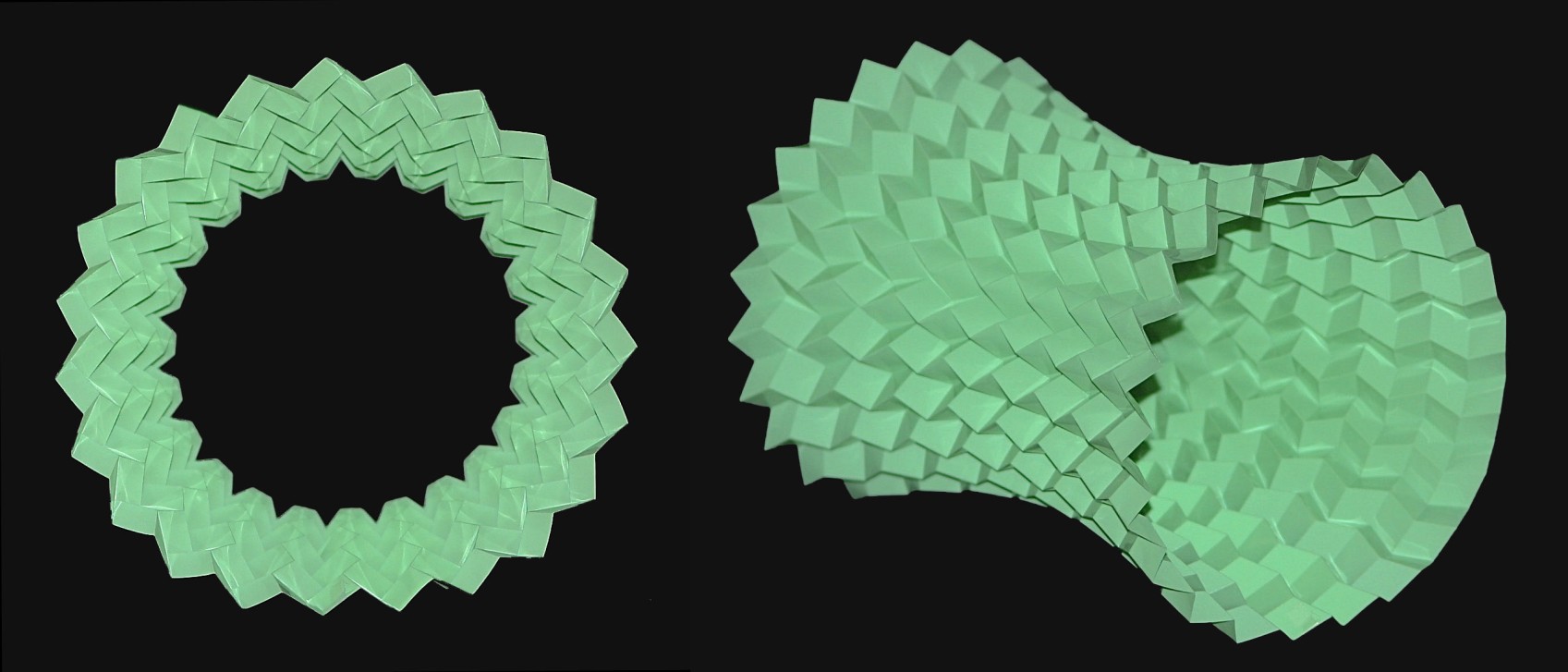}
    \caption{Two views of an origami-folded hyperboloid: A periodic ``Mars'' pattern is laser-etched into a PET sheet then hand-folded, circularly bent and fixated with adhesive tape. The pattern is auxetic but bends into a saddle. The letter explains the observed self-equilibrium geometry.}
    \label{fig:models}
\end{figure}

Morphing in origami metamaterials can be triggered in response to different loading conditions such as bending, twisting and stretching all of which couple in a manner unparalleled in other plates and shells; see Fig.~\ref{fig:models}. For instance, standard pure bending kinematics of plates imply that the curvatures in the bending plane and in the plane orthogonal to it, $\kappa_1$ and $\kappa_2$, are in a proportion equal and opposite to the in-plane Poisson's ratio $\nu$, i.e., $\nu=-\kappa_2/\kappa_1$. Thus auxetic plates tend to bend synclastically while anauxetic plates tend to bend anticlastically \cite{Naboni2016}. Origami metamaterials do not \cite{Schenk2011}. In fact, in what follows, we demonstrate that they typically exhibit the exact opposite behavior, i.e., $\nu=+\kappa_2/\kappa_1$. Other loads that induce morphing, such as ``pinching'', are unique to origami \cite{Grey2019}. Indeed, in plates and membranes, a pinch usually produces highly localized and singular deformations. By contrast, pinching an origami metamaterial brings on global and nearly-uniform deformations \cite{Seppecher2011, Filipov2015}. The consequence, as we shall prove, is that membrane energy penalizes, not strains, but their gradients, i.e., the Christoffel symbols~\cite{Mindlin1964, Gray2006, ciarlet2006}.

\begin{figure*}[ht!]
    \centering
    \includegraphics[width=\linewidth]{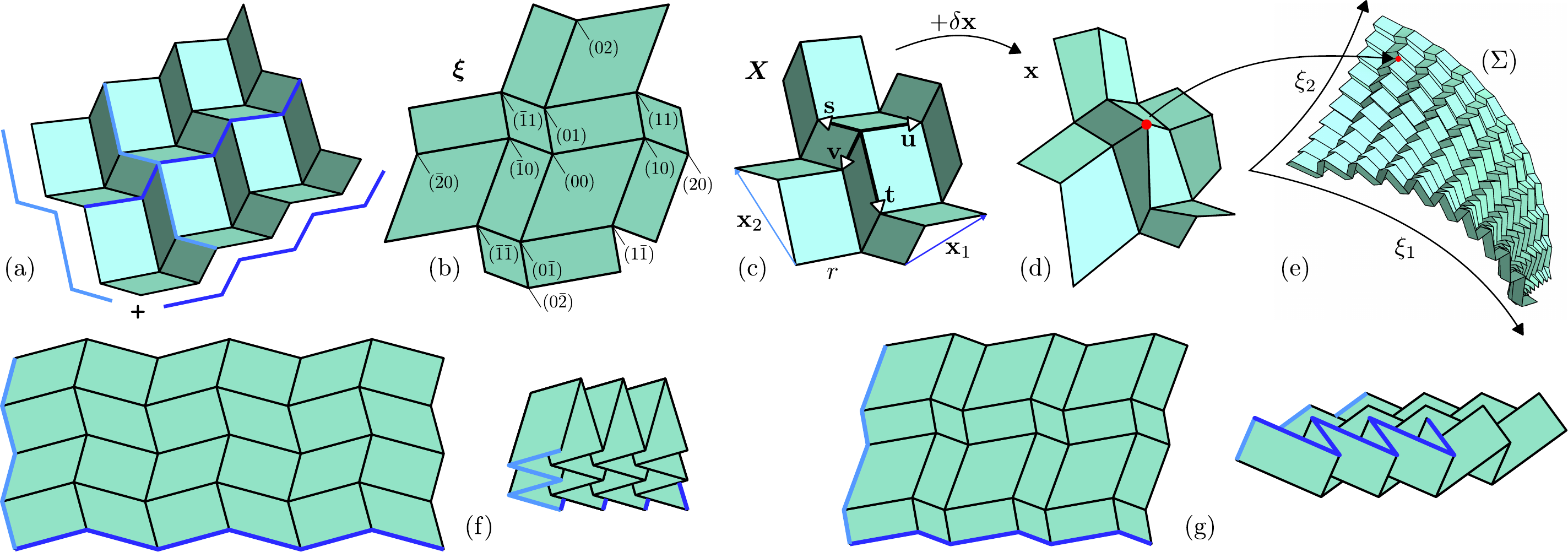}
    \caption{Zigzag sums. (a) A zigzag sum in a periodic reference configuration: translating one zigzag along another sweeps a polyhedral surface corresponding to a crease pattern complete with a natural assignment of mountain and valley folds. (b) A super-cell in an abstract reference state: indices $(ij)$ are discrete coordinates for the vertices; the central node is at $\gt\xi\equiv\gt\xi^{(00)}$. (c) A super-cell in a physical folded state: vectors $(\t u,\t v, \t s,\t t)$ are $\gt\xi$-dependent non-dimensional vectors aligned with the creases; they fully characterize fold-only configurations as well as the tangent vectors $\t x_1$ and $\t x_2$. (d) A super-cell following an infinitesimal bending $\delta\t x$ of the panels: note how the parallelograms are slightly distorted now. (e) A global view of a finitely deformed zigzag sum $(\Sigma)$: each super cell is designated by its central node's coordinates $\gt\xi=(\xi_1,\xi_2)$ (red dot) and is obtained by the process illustrated in panels (b-d). (f) An orthogonal zigzag sum: initially orthogonal zigzag directions remain orthogonal during folding ($g_{12}\equiv 0$). (g) An oblique zigzag sum: initially non-orthogonal zigzag directions are sheared during folding; even then, $g_{12}$ remains constant.}
    \label{fig:zigzags}
\end{figure*}

Previous contributions have succeeded in describing several aspects of origami mechanics, including calculations of Poisson's ratios and bending moduli \cite{Wei2013, Schenk2013, Pratapa2019}. Hereafter, we derive an effective medium theory complete with its field equations of compatibility and equilibrium. The theory naturally accounts for the observations described above and shows how the different deformation modes couple kinematically and energetically. The subject of the theory is a class of origami patterns that we call ``zigzag sums''. These are generated by translating one zigzag line along another so as to sweep a periodic polyhedral surface $\Sigma$ whose edges and panels we interpret as an origami crease pattern; see Fig.
~\ref{fig:zigzags}a. Equivalently, $\Sigma$ is a periodic crease pattern with a unit cell composed of four parallelograms. Four quadrilaterals per unit cell is as simple as origami tessellations can be without being trivial; moreover, here, the quads are parallelograms. Nonetheless, zigzag sums span various well-known patterns (e.g., the Miura-ori, the eggbox, ``Sakoda's staircase'' and ``Barreto's Mars'' patterns \cite{Lang2018}) and encompass their oblique, non-developable and non-flat-foldable variants by maintaining two basic requirements: periodicity and parallelism. Meanwhile, the zigzag generators provide a handful of geometric seeds that can readily tune their deformation paths as well as the underlying elastic behavior.

\begin{figure*}[ht!]
    \centering
    \includegraphics[width=\linewidth]{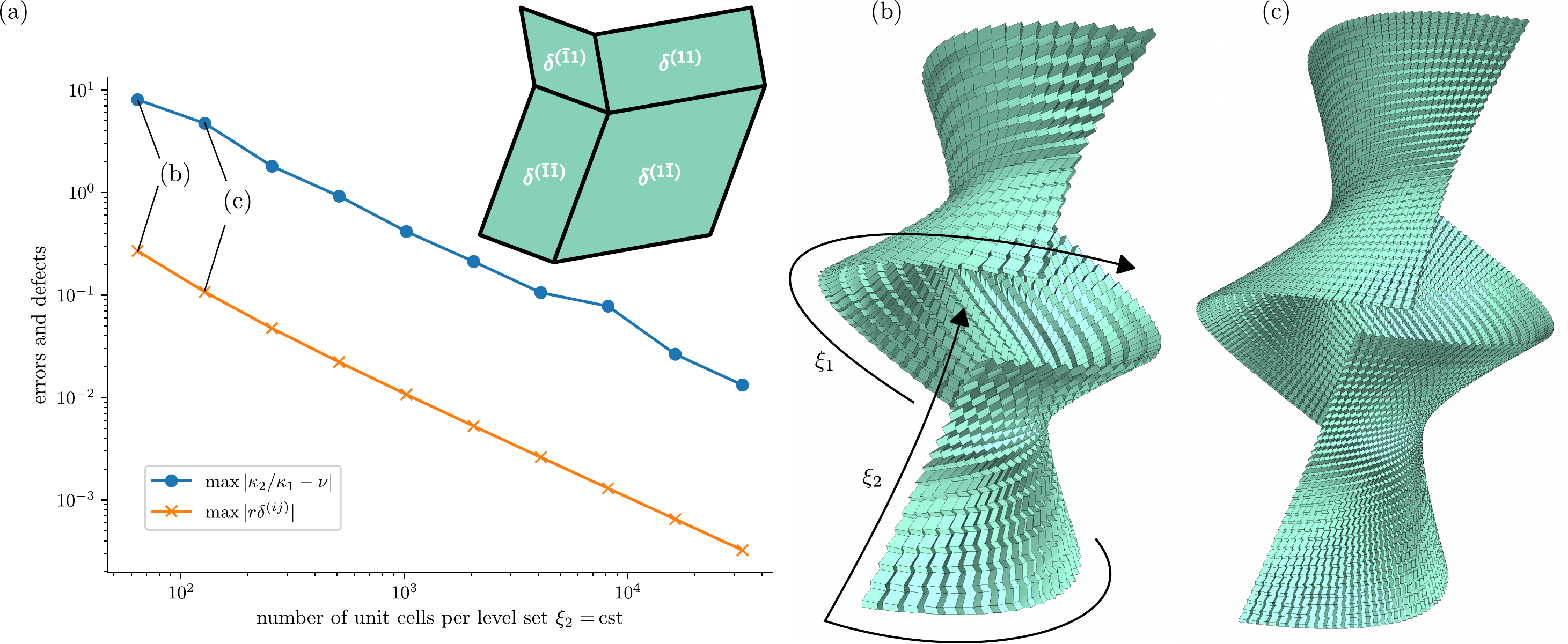}
    \caption{Convergence analysis. (a) Convergence speed: the maximum error in equation~\eqref{eq:Poisson} decreases like $r$ or, equivalently, like the reciprocal of the number of unit cells per row. The plot also shows that the maximum angular deflection $r\delta^{(ij)}$ decreases in the same fashion confirming the fold-to-bend orders of magnitude adopted in~\eqref{eq:scaling} and that, in the limit $r\to 0$, all panels are flat. The inset shows the unit cell adopted for this example. (b,c) Two iterates: snapshots of the way in which the zigzag sum converges to a smooth surface taken for $64$ and $128$ unit cells per row. Note that the zigzag sum is finitely and non-uniformly stretched and bent. Illustrated surfaces are constructed by ``propagating'' inextensibility constraints away from a given boundary; see~\cite{Nassar2018b} for details.}
    \label{fig:convergence}
\end{figure*}

\section{Geometric mechanics of zigzag sums}
The deformations of a zigzag sum $\Sigma$ are the outcome of two competing mechanisms \cite{Lechenault2014}. The first is the folding and unfolding of $\Sigma$ along its edges. The second is the isometric bending of its panels. Signature origami kinematics emerges when the bending stiffness of the panels is significantly larger than the folding stiffness of the edges.
Accordingly, we can assume that individual panels bend infinitesimally; $\Sigma$, on the other hand, can still embrace finitely curved surfaces by accumulating small deflections across many unit cells. Formally, we characterize the deformations of $\Sigma$ in terms of a continuous parametrization $\t x=\t x(\gt\xi)$ of its mid-plane, where $\t x$ is the current position of vertex $\gt\xi$, \highlight{with $\gt\xi=(\xi_1,\xi_2)$ being the effective curvilinear coordinate along the zigzags directions}; see Fig.~\ref{fig:zigzags}b-e. Consider then a non-elementary unit cell, i.e., a supercell, centered about vertex $\gt\xi$ and containing the neighboring vertices $\gt\xi^{(ij)}$ placed at $\t x^{(ij)}$. Then, the adopted origami kinematics imply that position
\[\label{eq:scaling}
    \t x^{(ij)} = \t X^{(ij)} + r^2\delta\t x^{(ij)}
\]
is attained by a fold-only motion $\gt\xi^{(ij)}\mapsto\t X^{(ij)}$ (Fig.~\ref{fig:zigzags}b-c) followed by a bending-induced perturbation $r^2\delta\t x^{(ij)}$ that is small compared to the characteristic size of the panels denoted $r$ (Fig.~\ref{fig:zigzags}c-d). In the limit of a tessellation where $\Sigma$ contains a large number of unit cells, i.e., for $r\to 0$, the plane locally tangent to $\Sigma$ is spanned by two vectors
\[
\begin{split}
\t x_1\equiv\frac{\partial\t x}{\partial\xi_1} &= \lim_{r\to 0}\frac{\t x^{(10)}-\t x^{(\bar10)}}{r}=\t u-\t v,\\
\t x_2\equiv \frac{\partial\t x}{\partial\xi_2} &= \lim_{r\to 0}\frac{\t x^{(01)}-\t x^{(0\bar1)}}{r}=\t s-\t t,
\end{split}
\]
where $\t u$, $\t v$, $\t s$ and $\t t$ are the crease vectors meeting at vertex $\gt\xi$ and normalized with respect to $r$. The apparent in-plane deformations of $\Sigma$ can therefore be quantified in terms of an effective metric tensor $\t g$ of components
\[\label{eq:constr1}
g_{\mu\nu}=\scalar{\t x_\mu,\t x_\nu} = \begin{bmatrix}
\norm{\t u-\t v}^2 & \scalar{\t u-\t v,\t s-\t t} \\
\scalar{\t u-\t v,\t s-\t t} & \norm{\t s-\t t}^2
\end{bmatrix}.
\]
Note that bending-related contributions disappear in the limit $r\to 0$ and can be disregarded for now.

It is noteworthy that $g_{12}=\scalar{\t u,\t s} + \scalar{\t v,\t t} - \scalar{\t u,\t t}-\scalar{\t v,\t s}$ is a motion constant (i.e., $\dd g_{12}\equiv 0$) as it is equal to a combination of constant lengths and angles. Thus, if $g_{12}=0$ so that the zigzags $\xi_1$- and $\xi_2$-contours are initially orthogonal, then orthogonality is maintained in any subsequent motion (Fig.~\ref{fig:zigzags}f). However, in oblique tessellations, with $g_{12} = \norm{\t x_1}\norm{\t x_2}\cos\theta \neq 0$, stretching couples to shearing according to
\[
    \tan\theta\,\dd\theta =
    \frac{\dd\!\norm{\t x_1}}{\norm{\t x_1}}
    +
    \frac{\dd\!\norm{\t x_2}}{\norm{\t x_2}}
    =\frac{\dd g}{2g_{11}g_{22}}
\]
with $g=\det\t g$. Rather intuitively then, shearing, understood as aligning the zigzags ($\theta\to 0$ or $\pi$), always goes to reduce the pattern's effective area (Fig~\ref{fig:zigzags}g). As for the stretch ratios in directions $1$ and $2$, they are related to one another through a geometric Poisson's ratio $\nu\equiv
    -g_{11}/g_{22}\times\dd g_{22}/\dd g_{11}$.
A general expression for $\nu$ is pursued in the Supplemental Material \cite[App. A]{note}.

Most important is the observation that the fold-only motion $\gt\xi^{(ij)}\mapsto\t X^{(ij)}$ maintains crease parallelism and periodicity: this is truly where the fact that $\Sigma$ is made out of parallelograms is crucial. Hence, folding alone produces no curvature and, in that regard, it is bending that takes over now. This is most apparent in the expressions of the second-order derivatives which, in the limit $r\to 0$, read
\[
\begin{split}
\t x_{11}&=\delta\t x^{(20)}-2\delta\t x+\delta\t x^{(\bar{2}0)},\\
\t x_{22}&=\delta\t x^{(02)}-2\delta\t x+\delta\t x^{(0\bar{2})},\\
\t x_{12}&=\delta\t x^{(11)}-\delta\t x^{(\bar{1}1)}-\delta\t x^{(1\bar{1})}+\delta\t x^{(\bar{1}\bar{1})},
\end{split}
\]
where $\t x_{\mu\nu}\equiv\partial^2\t x/\partial\xi_\mu\partial\xi_\nu$. Therefore, the $\t x_{\mu\nu}$ at $\gt\xi$ are linear functions of the bending DOFs of the super cell centered at $\gt\xi$. \highlight{That being said, not all $\delta\t x^{(ij)}$ are compatible with the inextensibility constraints. Ultimately, it is possible to show that $(\t x_{11}, \t x_{22}, \t x_{12})$ belong to a linear subspace of $\R^3\times\R^3\times\R^3$ spanned by 4 DOFs, each attributed to a planarity defect $(\delta^{(ij)})$ of one of the four panels of the central unit cell (Fig~\ref{fig:convergence}a) \cite[App. B]{note}.} By the rank-nullity theorem, the $\t x_{\mu\nu}$ satisfy $9-4=5$ linear compatibility equations. Straightforward calculations show that the first two of these equations are $g_{12,1}=g_{12,2}=0$, i.e., they re-produce $\dd g_{12}=0$. The other three relations are far more interesting; they combine into the vector identity \cite[App. B]{note}
\[\label{eq:identity}
\t x_{22} = \frac{\scalar{\t u,\t t\wedge\t s}\scalar{\t v,\t t\wedge\t s}}
{\scalar{\t t,\t u\wedge\t v}\scalar{\t s,\t u\wedge\t v}}\t x_{11}.
\]
The tangent components of $\t x_{22}$ and $\t x_{11}$ describe the strain gradients in the mid-plane of $\Sigma$ in direction $\t x_1$ and $\t x_2$ whereas the normal components describe the normal curvatures $\kappa_1$ and $\kappa_2$ of the mid-plane in the same directions. The above identity states that these respective quantities necessarily occur in equal proportions. Specifically, the in-plane and out-of-plane Poisson's ratios, $\nu$ and $-\kappa_2/\kappa_1$, of a zigzag sum $\Sigma$ are equal and opposite and are given by the current geometry of the folds as per
\[\label{eq:Poisson}
\nu = \frac{\kappa_2}{\kappa_1} = \frac{g_{11}}{g_{22}}\frac{\scalar{\t u,\t t\wedge\t s}\scalar{\t v,\t t\wedge\t s}}
{\scalar{\t t,\t u\wedge\t v}\scalar{\t s,\t u\wedge\t v}},
\]
with $\kappa_\mu=\scalar{\t x_{\mu\mu},\t n}/g_{\mu\mu}$ (no sum), $\t n = \t x_1\wedge\t x_2/\norm{\t x_1\wedge\t x_2}$ being the unit normal \cite[App. C]{note}.

Identity~\eqref{eq:Poisson} is of an asymptotic nature: in the limit $r\to 0$ of infinitesimal unit cells, the Poisson's ratios of a zigzag sum $\Sigma$ converge to one another. In other, more practical, terms, the equality is accurate within an error proportional to $r$. See Fig~\ref{fig:convergence} for a numerical illustration of the convergence process and the relevant error bounds. Identity~\eqref{eq:Poisson} is also universal: it holds whether $\Sigma$ is developable, flat-foldable, symmetric, rectangular, or not. Therein, the mixed products (e.g., $\scalar{\t u,\t t\wedge\t s}$) have their signs determined by the mountain-valley assignment of the concerned folds. This allows us to deduce the rather nice corollary: if a zigzag sum $\Sigma$ has an odd number of mountain folds meeting at a vertex then: $(i)$ it is necessarily auxetic; and $(ii)$ it necessarily bends into a saddle. In particular, all developable zigzag sums are auxetic and bend into saddles. Note that these results are only valid for nondegenerate configurations, i.e., where all folds are partially folded. Indeed, when a fold is flat, be it open or closed, mountain and valley assignments coalesce and the Poisson's ratio could become null, infinite or multivalued \cite{Pratapa2019}. Note also that the proven identity features the elongations and curvatures in two specific material directions $\xi_1$- and $\xi_2$-contours, be them orthogonal or not; in particular, the identity bears no immediate consequences on the Poisson's coefficient defined for two general orthogonal directions.

\begin{figure*}[ht!]
    \centering
    \includegraphics[width=\linewidth]{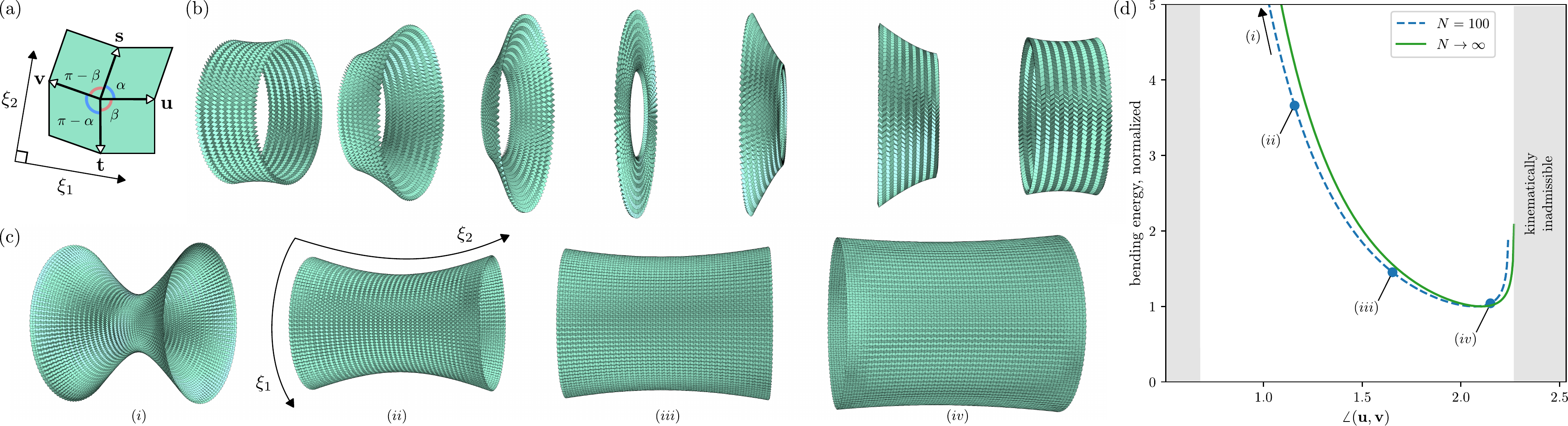}
    \caption{``Mars''-folded origami pillars are one-sheeted hyperboloids. (a) A unit cell of the ``Mars'' pattern: the supplementary opposite angles ensure developability and flat-foldability; an equilateral pattern is further rectangular. (b) Action of the first DOF on the shape of the pillar: given a small enough aspect ratio $L_2/L_1$, the pillar can be inverted inside-out; here, the pattern has $N=100$ unit cells per meridian, and $8$ cells per parallel, with a flat-unfolded aspect ratio $L_2/L_1= 8\%$. (c) Action of the second DOF: the pattern flattens as the opening angle $\angle(\t u,\t v)$ at a given parallel is increased; here, the pattern has $N=100$ unit cells per meridian and $40$ unit cells per parallel with $L_2/L_1= 40\%$. (d) Bending energy: plot of normalized total bending energy as a function of the opening angle $\angle(\t u,\t v)$ at the equator for origami pillars centered about the equator with $L_2/L_1=40\%$; the discretely summed energy for $N=100$ is plotted against the continuum limit $N\to\infty$; data points $(i-iv)$ are the ones illustrated on panel (c); greyed out domain cannot be accessed with the current aspect ratio; the continuum theory satisfactorily predicts both energy and kinematical admissibility. Adopted values: $\alpha=\pi/3$, $\beta=\pi/2$.}
    \label{fig:hyperboloids}
\end{figure*}

\section{Self-equilibrium of origami pillars}
The equality of the Poisson's ratios greatly constrains the 3D geometries accessible to a zigzag sum $\Sigma$ \cite{Nassar2017a, Nassar2017e, Nassar2018b, zheng2021continuum}. Consider, for tractability, the ``Mars'' tessellation of Fig.~\ref{fig:hyperboloids}a. The tessellation lacks mirror symmetry but is otherwise developable, flat-foldable and equilateral. We find that compatible metrics and curvatures satisfy \cite[App. D]{note}
\[\label{eq:metricEx}
\begin{gathered}
g_{12}=0,\quad
g_{11}(4-g_{22}) = 
4(\cos\alpha+\cos\beta)^2,\\
\frac{\kappa_2}{\kappa_1} = \nu =
-4\frac{(\cos\alpha+\cos\beta)^2}{g_{11}g_{22}}.
\end{gathered}
\]
This is in fact a system of non-linear PDEs weighing on the admissible configurations $\t x$. For instance, we explore the configurations of a rectangular domain of flat dimensions $L_1\times L_2$ that are folded and wrapped around an axis of symmetry into an origami pillar \cite{Lang2018}. These pillars are surfaces of revolution with a Cartesian parametrization of the form
\[
\t x(\gt\xi)=
(\rho(\xi_2)\cos(\omega\xi_1),\rho(\xi_2)\sin(\omega\xi_1),z(\xi_2)).
\]
Here, the axis of symmetry is the $z$-axis, $\xi_1$-contours are meridian lines and $\xi_2$-contours are parallel lines. Equations~\eqref{eq:metricEx} then simplify into a system of ODEs governing $z$ and $\rho$. Its solutions constitute a 2-DOF family of one-sheeted hyperboloids (Fig.~\ref{fig:hyperboloids}b,c) \cite[App. E]{note}. The range of motion of each pillar is bounded by the maximally folded and unfolded states of the pattern. Thus, for increasing aspect ratio $L_2/L_1$, the pillar will have a smaller range of motion, until it jams, or even becomes impossible to form without tearing.

Among all accessible pillars, the ones that are in self-equilibrium exhibit minimum levels of strain energy $\psi$. Strain energy is composed of two contributions: crease folding and panel bending. Crease folding energy takes the form of a classical, however nonlinear, membrane energy. In what follows, we neglect this contribution and focus on the less-explored influence of panel bending \cite{Grey2020}. For ``Mars''-folded origami pillars, centered about the equator, strain energy is exemplified on Fig.~\ref{fig:hyperboloids}d. It appears that unfolding the pattern, by increasing the angle $\angle(\t u,\t v)$, reduces bending energy. The unfolding flattens the pillar and reduces both curvatures $\kappa_1$ and $\kappa_2$. That trend continues up to a point where energy is minimum (near state $(iv)$) and beyond which any further unfolding, or flattening, of the pattern actually causes a dramatic increase in bending energy. This suggests that bending energy $\psi$ depends on some other deformation measures besides curvatures. As a matter of fact, the bending energy density $b$, where $\psi=\int_\Omega b\dd\gt\xi$, $\Omega$ being the reference domain of the pattern, is a quadratic form of the planarity defects $\delta^{(ij)}$ of the panels. For their part, the defects are linear forms of the full, both out-of-plane and in-plane, components of the parametrization's second derivatives $\t x_{\mu\nu}$. The out-of-plane components $\Gamma_{3\mu\nu}\equiv\scalar{\t n,\t x_{\mu\nu}}$ are the coefficients of the second fundamental form; they quantify the curvatures and torsion of the embraced surface. The in-plane components $\Gamma_{\sigma\mu\nu}\equiv\scalar{\t x_\sigma,\t x_{\mu\nu}}$ are the Christoffel symbols; they quantify, not the strains, but their gradient, i.e., terms of the form $g_{\mu\nu,\sigma}$. Accordingly, $b=b(\gt\Gamma;\t g)$ is a metric-dependent quadratic form of the curvatures, torsion, and strain gradient. Detailed derivations of the expression of $b$ leading to Fig.~\ref{fig:hyperboloids}d can be found in the Supplemental Material \cite[App. F, G]{note}.

Back to the pillars, unfolding the pattern reduces both curvatures and strain-gradients. However, as the geometric rigidities associated with the curvatures (i.e., $\partial^2b/\partial\Gamma_{3\mu\nu}$) remain bounded, the ones associated with the strain gradient (i.e., $\partial^2b/\partial\Gamma_{\sigma\mu\nu}$) diverge for states that are close to being flat. Such states occur near the outer rims of sufficiently unfolded pillars; their presence further signals that the pillar has reached the boundary of the kinematically admissible domain, hence the energy blow-up observed on Fig.~\ref{fig:hyperboloids}d. Strain-gradient energy further dominates the response of any plane, non-uniform, state such as the ``ring'' observed midway through Fig.~\ref{fig:hyperboloids}b. Indeed, uniformly folded plane states do not engage panel bending and therefore have zero energy. By contrast, non-uniformly folded states, with gradients of folding angles, cannot be achieved without panel bending. When the state is plane, it has zero curvatures and torsion, and energy becomes function of the strain gradient alone \cite{Seppecher2011}. More generally, it is noteworthy that $\psi$ does not penalize strains or fold angles, however large, so long as they are uniform. In fact, $\psi$ does not even refer to a specific natural state or any specific natural fold angles in reference to which strains should be measured \cite{ciarlet2006}. Instead, it refers to a higher-order strain measure, namely $\gt\Gamma$, which quantifies not how much the pattern is folded, but rather how far it is from being uniformly folded.

\section{Conclusion}
In conclusion, we proposed a continuum theory of origami-folded shells where the crease pattern is a zigzag sum. The theory accounts for geometric nonlinearities and accurately predicts the states that are accessible under isometric folding thanks to a universal identity relating in-plane deformations and out-of-plane curvatures. The study further demonstrates that the energy density of origami-folded shells can heavily depend on the in-plane strain-gradient in a way that is unparalleled in classical shell theories. \highlight{It should be possible to extend the proposed theory, in one form or another, to more general tessellations that possess periodic folding motions. By contrast, alternative approaches are likely to be needed to tackle tessellations that do not fold periodically (e.g., Huffman grids and Yoshimura-like patterns  \cite{Evans2015a})}. Finally, it would be of interest to investigate how the proposed theory can inform shape programming and morphing planning of origami and kirigami structures \cite{tachi2013, Dudte2016a, Guseinov2020, Siefert2020, Jin2020, Feng2020, Li2021,  Yang2021, Melancon2021a} for robotics and 4D printing applications.

\section*{Acknowledgements}
HN acknowledges support by the NSF under CAREER award No. CMMI-2045881. AL acknowledges the support of the French Agence Nationale de la Recherche (ANR), under grant No. ANR-17-CE08-0039 (project ArchiMatHOS). EW acknowledges support by the NASA Missouri Space Grant Consortium.


\appendix
\section{The in-plane Poisson's ratio: General case}
The metric $\t g$ of a zigzag sum has the components $g_{11}=\norm{\t u-\t v}^2$, $g_{22}=\norm{\t s-\t t}^2$ and $g_{12}=\scalar{\t u-\t v,\t s-\t t}$. Note that the scalar products $\scalar{\t u,\t s}$, $\scalar{\t u,\t t}$, $\scalar{\t v,\t s}$ and $\scalar{\t v,\t t}$ are combinations of lengths and angles that are isometrically preserved; they are motion constants and so is $g_{12}$. It is somewhat more challenging to derive a relationship between $g_{11}$ and $g_{22}$; this is done next. First, note that $\t s$ and $\t t$ can be decomposed into
\[
\begin{split}
    \t s &= \scalar{\t s,\t u}\t u^*+
    \scalar{\t s,\t v}\t v^*+
    s_w \t w,\\
    \t t &= \scalar{\t t,\t u}\t u^*+
    \scalar{\t t,\t v}\t v^*+
    t_w \t w,
\end{split}
\]
with
\[
    \t u^* = \frac{\t v\wedge\t w}{\norm{\t u\wedge\t v}},\quad
    \t v^* = \frac{\t w\wedge\t u}{\norm{\t u\wedge\t v}},\quad
    \t w = \frac{\t u\wedge\t v}{\norm{\t u\wedge\t v}}.
\]
Components $s_w$ and $t_w$ are easily determined by considering the magnitudes of $\t s$ and $\t t$. Indeed, we have
\[\label{eq:swtw}
    \begin{split}
        s_w^2 = s^2 - \frac{\scalar{\t s,\t u}^2v^2+\scalar{\t s,\t v}^2u^2-2\scalar{\t s,\t u}\scalar{\t s,\t v}\scalar{\t u,\t v}}{\norm{\t u\wedge\t v}^2},\\
        t_w^2 = t^2 - \frac{\scalar{\t t,\t u}^2v^2+\scalar{\t t,\t v}^2u^2-2\scalar{\t t,\t u}\scalar{\t t,\t v}\scalar{\t u,\t v}}{\norm{\t u\wedge\t v}^2},
    \end{split}
\]
with $u=\norm{\t u}$, and so on. Therein, note that the only variables are
\[
    \scalar{\t u,\t v} = \frac{u^2+v^2-g_{11}}{2},\quad
    \norm{\t u\wedge\t v}^2 = u^2v^2-\scalar{\t u,\t v}^2,
\]
and are both functions of $g_{11}$. Last, we write
\begin{multline}
\label{eq:g11g22}
g_{22} = 
\frac{\scalar{\t s-\t t,\t u}^2v^2+\scalar{\t s-\t t,\t v}^2u^2}{\norm{\t u\wedge\t v}^2}\\
-
\frac{2\scalar{\t s-\t t,\t u}\scalar{\t s-\t t,\t v}\scalar{\t u,\t v}}{\norm{\t u\wedge\t v}^2}\\
+ (s_w-t_w)^2,
\end{multline}
and accordingly deduce $g_{22}$ as a function of $g_{11}$, as well as $\nu$ as a function of $g_{11}$. It is worth mentioning that $g_{11}$ and $g_{22}$ have upper and lower bounds corresponding to some creases being maximally folded or unfolded. Beyond these bounds the pattern would penetrate itself or break apart. These situations can be avoided by enforcing $s_w^2>0$ and $t_w^2>0$.
\section{The admissible second-order derivatives}
Consider the vertices of a super cell placed at $\t x^{(ij)}=\t X^{(ij)}+r^2\delta\t x^{(ij)}$ where the $\t X^{(ij)}$ describe a pre-folded state and the $r^2\delta\t x^{(ij)}$ are bending-induced perturbations. Similarly, the central vertex $\gt\xi$ is at $\t X$ subsequent to the fold-only motion and is perturbed by $r^2\delta\t x$ subsequent to the infinitesimal bending of the panels. Note that the second-order derivatives $\t x_{\mu\nu}$ are invariant by composition with linear motions including rigid body motions and periodic stretching and contraction. Accordingly, it is possible, without loss of generality, to default the displacements of the central four creases to zero. In other words, we set
\[\label{eq:defZeroDisp}
\delta\t x=\delta\t x^{(10)}
=\delta\t x^{(01)}
=\delta\t x^{(\bar 10)}
=\delta\t x^{(0\bar 1)}
=\t 0.
\]
Furthermore, by linearity, the contributions of the remaining displacements can be investigated independently then superposed.

So let $\delta\t x^{(\bar11)}$, $\delta\t x^{(\bar1\bar1)}$, $\delta\t x^{(1\bar1)}$, $\delta\t x^{(\bar20)}$ and $\delta\t x^{(0\bar2)}$ all be null for now and consider a nonzero displacement $\delta\t x^{(11)}$: the latter must be orthogonal to $\t u$ so as to preserve the length of $\t x^{(11)}-\t x^{(01)}$ as well as to $\t s$ so as to preserve the length of $\t x^{(11)}-\t x^{(10)}$, to leading order. Hence,
\[
\delta\t x^{(11)} = \delta^{(11)}\t u\wedge\t s
\]
for some planarity defect $\delta^{(11)}$. Similarly, we have
\[
\delta\t x^{(02)} = \delta^+\t v\wedge\t t,\quad
\delta\t x^{(20)} = \delta^-\t v\wedge\t t.
\]
Expressions for $\delta^\pm$ are easily obtained by observing that the lengths of $\t x^{(11)}-\t x^{(02)}$ and $\t x^{(11)}-\t x^{(20)}$ are preserved. Indeed, infinitesimal inextensibility implies
\[\label{eq:meth1}
\begin{split}
  \scalar{\t t+\t u,\delta^{(11)}\t u\wedge\t s-\delta^+\t v\wedge\t t}&=0,\\
  \scalar{\t v+\t s,\delta^{(11)}\t u\wedge\t s-\delta^-\t v\wedge\t t}&=0,
\end{split}
\]
namely,
\[\label{eq:meth2}
\frac{\delta^+}{\delta^-} = 
\frac{\scalar{\t u\wedge\t s,\t t}
\scalar{\t v\wedge\t t,\t s}}
{\scalar{\t v\wedge\t t,\t u}
\scalar{\t u\wedge\t s,\t v}}.
\]
As for the $\t x_{\mu\nu}$, they are given by
\[\label{eq:xmunuelem}
\begin{split}
    \t x_{11}&={\delta\t x^{(20)}}={\delta^-}\t v\wedge\t t,\\
    \t x_{22}&={\delta\t x^{(02)}}={\delta^+}\t v\wedge\t t,\\
    \t x_{12}&={\delta\t x^{(11)}}=\delta^{(11)}\t u\wedge\t s.
\end{split}
\]
In particular,
\[
\t x_{22} = 
\frac{\scalar{\t u\wedge\t s,\t t}
\scalar{\t v\wedge\t t,\t s}}
{\scalar{\t v\wedge\t t,\t u}
\scalar{\t u\wedge\t s,\t v}}
\t x_{11}.
\]
The contributions of $\delta\t x^{(\bar11)}$, $\delta\t x^{(\bar1\bar1)}$ and $\delta\t x^{(1\bar1)}$ (i.e., $\delta^{(\bar11)}$, $\delta^{(\bar1\bar1)}$ and $\delta^{(1\bar1)}$, respectively) can be readily calculated in the same fashion and shown to yield exactly the same constraint on $\t x_{22}$ and $\t x_{11}$. This can be predicted by appreciating the symmetric way in which the proportionality coefficient depends upon the crease vectors. In conclusion, the above constraint holds for any combination of admissible infinitesimal displacements of the super cell.

Last, note that
\[
\begin{split}
\scalar{\t x_{12},\t x_1}&=
\scalar{\delta^{(11)}\t u\wedge\t s,\t u-\t v}\\
&=-\delta^{(11)}\scalar{\t u\wedge\t s,\t v}\\
&=
-\delta^-\scalar{\t v\wedge\t t,\t s}\\
&=
-\delta^-\scalar{\t v\wedge\t t,\t s-\t t}\\
&=
-\scalar{\t x_{11},\t x_2}.
\end{split}
\]
That is: $\partial_1\!\scalar{\t x_1,\t x_2}=0$. In the same manner, one shows that $\partial_2\!\scalar{\t x_1,\t x_2}=0$ in order to recover $\dd g_{12}=0$ as claimed in the main text.
\section{The in- and out-of-plane Poisson's coefficients are equal and opposite}
Projecting the proportionality constraint of $\t x_{11}$ and $\t x_{22}$ along the normal $\t n$ yields
\[
\kappa_2g_{22} = 
\frac{\scalar{\t u\wedge\t s,\t t}
\scalar{\t v\wedge\t t,\t s}}
{\scalar{\t v\wedge\t t,\t u}
\scalar{\t u\wedge\t s,\t v}}
\kappa_1g_{11}
\]
by the definition of the normal curvatures in directions 1 and 2. Projecting along $\t x_1$ yields
\[
\scalar{\t x_{22},\t x_1}= 
\frac{\scalar{\t u\wedge\t s,\t t}
\scalar{\t v\wedge\t t,\t s}}
{\scalar{\t v\wedge\t t,\t u}
\scalar{\t u\wedge\t s,\t v}}
\scalar{\t x_{11},\t x_1}.
\]
But
\[
\begin{split}
\scalar{\t x_{22},\t x_1}&=
\partial_2\!\scalar{\t x_2,\t x_1}
-\scalar{\t x_2,\t x_{12}}\\
&=
\partial_2\!\scalar{\t x_2,\t x_1}
-\frac12\partial_1\!\scalar{\t x_2,\t x_2}\\
&=
-\frac12\partial_1g_{22},
\end{split}
\]
since $\dd g_{12}=0$. Also, $\scalar{\t x_{11},\t x_1}=\partial_1g_{11}/2$. Projecting over $\t x_2$ yields similar relations so that overall
\[
-\frac{\dd g_{22}}{2}= 
\frac{\scalar{\t u\wedge\t s,\t t}
\scalar{\t v\wedge\t t,\t s}}
{\scalar{\t v\wedge\t t,\t u}
\scalar{\t u\wedge\t s,\t v}}
\frac{\dd g_{11}}{2}.
\]
Hence, the in-plane Poisson's coefficient is
\[
\nu = -\frac{\dd g_{22}/g_{22}}{\dd g_{11}/g_{11}}
= 
\frac{g_{11}}{g_{22}}
\frac{\scalar{\t u\wedge\t s,\t t}
\scalar{\t v\wedge\t t,\t s}}
{\scalar{\t v\wedge\t t,\t u}
\scalar{\t u\wedge\t s,\t v}}
=
\frac{\kappa_2}{\kappa_1}
\]
and matches, up to a sign, the out-of-plane Poisson's coefficient $-\kappa_2/\kappa_1$.
\section{The Poisson's ratios of the ``Mars'' pattern}
The ``Mars'' pattern is a developable flat-foldable zigzag sum. Let us suppose that the pattern is equilateral so that $u=v=s=t$. Now we have
\[
    \scalar{\t u,\t s}=-\scalar{\t v,\t t} = \cos\alpha,\quad
    \scalar{\t u,\t t}=-\scalar{\t v,\t s} = \cos\beta.
\]
We are at liberty to adopt a mountain-valley assignment convention: we let creases $\t s$, $\t t$ and $\t v$ be mountains and let crease $\t u$ be a valley. Then, inspection of~\eqref{eq:swtw} reveals that $s_w=-t_w$. This allows to greatly simplify~\eqref{eq:g11g22} ultimately into
\[
g_{22} = 
4-4\frac{(\cos\alpha+\cos\beta)^2}{g_{11}}.
\]
The Poisson's coefficient is deduced by differentiation. It reads
\[
    \nu = -4\frac{(\cos\alpha+\cos\beta)^2}{g_{11}g_{22}}
\]
and is automatically equal to $\kappa_2/\kappa_1$.

Recall that $g_{11}$ and $g_{22}$ are bounded. In the maximally unfolded state, $\cos(\t u,\t v)=-\cos(\alpha-\beta)$ so that, in any case, $g_{11}$ remains smaller than $2+2\cos(\alpha-\beta)$. The other bounds are similarly deduced and, overall, we have
\[\label{eq:metricBounds}
    \begin{split}
    4\cos^2\left(\frac{\alpha+\beta}{2}\right) &\leq g_{11} \leq 4\cos^2\left(\frac{\alpha-\beta}{2}\right),\\
    4\sin^2\left(\frac{\alpha-\beta}{2}\right) &\leq g_{22} \leq 4\sin^2\left(\frac{\alpha+\beta}{2}\right).
    \end{split}
\]
\section{``Mars''-folded hyperboloids}
Consider an axisymmetric origami pillar folded out of an equilateral ``Mars'' pattern with a parametrization $\t x\in\R^3$ of the form
\[
    \t x = (\rho(\xi_2)\cos(q\xi_1),
    \rho(\xi_2)\sin(q\xi_1),
    z(\xi_2)).
\]
We denote $L_1$ and $L_2$ the dimensions of the origami pattern in its flat, maximally unfolded, reference state. We then let $\xi_1\in[-S_1/2,S_1/2]$ and $\xi_2\in[-S_2/2,S_2/2]$ with
\[
S_1 = \frac{L_1}{2\cos((\alpha-\beta)/2)},\quad
   S_2 = \frac{L_2}{2\sin((\alpha+\beta)/2)}.
\]
The pillar is obtained by folding the pattern then glueing together the ends $\xi_1=0$ and $\xi_1=S_1$. In other words, we let $q=2\pi/S_1$.

With these notations, it straightforward to see that
\[
\begin{split}
    \t x_1 &= (-q\rho(\xi_2)\sin(q\xi_1),
    q\rho(\xi_2)\cos(q\xi_1),
    0),\\
    \t x_2 &= (\rho'(\xi_2)\cos(q\xi_1),
    \rho'(\xi_2)\sin(q\xi_1),
    z'(\xi_2)),\\
    \t x_{11} &= (-q^2\rho(\xi_2)\cos(q\xi_1),
    -q^2\rho(\xi_2)\sin(q\xi_1),
    0),\\
    \t x_{22} &= (\rho''(\xi_2)\cos(q\xi_1),
    \rho''(\xi_2)\sin(q\xi_1),
    z''(\xi_2)),
\end{split}
\]
with $\rho'\equiv\dd\rho/\dd\xi_2$, $\rho''\equiv\dd\rho'/\dd\xi_2$ and so on.

Then,
\[
    g_{11} = q^2\rho^2
    \quad\text{and}
    \quad
    g_{22} = \rho'^2+z'^2
\]
depend on one another through the in-plane kinematical constraint
\[
    \rho'^2 + z'^2 = 
    4 - 4\frac{(\cos\alpha+\cos\beta)^2}{q^2\rho^2}.
\]
The out-of-plane constraint, i.e., $\t x_{11}\parallel\t x_{22}$, immediately yields
\[
    z'' = 0.
\]
Together, these two ODEs can be integrated into
\[
    z' = \text{cst},\quad
    \rho = \sqrt{4c^2(\xi_2-\xi_o)^2 + \frac{(\cos\alpha+\cos\beta)^2}{q^2c^2}},
\]
with $c^2 = 1 - z'^2/4$. Hence, the origami pillar embraces a hyperboloid. The pattern does not cover the whole hyperboloid however and is limited to the band spanned by $\xi_2\in[-S_2/2,S_2/2]$. Overall then, and up to rigid body motions, the folded pillar has two degrees of freedom. The first is $c$: it describes the shape of the hyperboloid; as it goes from 1 to 0, the hyperboloid changes from a ``yoyo'' into a cylinder. The second is $\xi_o$: it specifies how the band of length $L_2$ that the pattern covers is centered on or offset away from the equator.

Not all values $(c,\xi_o)$ are admissible however since candidate solutions must further satisfy the upper and lower bounds weighing on the metric, namely~\eqref{eq:metricBounds}. In the present case, these read
\[
    4\cos^2((\alpha+\beta)/2) \leq q^2\rho^2 \leq 4\cos^2((\alpha-\beta)/2).
\]
For a centered band ($\xi_o=0$) in particular, these bounds reduce to
\[
\begin{split}
    4\cos^2\left(\frac{\alpha+\beta}{2}\right) &\leq \frac{(\cos\alpha+\cos\beta)^2}{c^2},  \\
    c^2q^2S_2^2 + \frac{(\cos\alpha+\cos\beta)^2}{c^2} &\leq
    4\cos^2\left(\frac{\alpha-\beta}{2}\right),
\end{split}
\]
and place a maximum bound on the aspect ratio $S_2/S_1=qS_2/(2\pi)$ of origami pillars that can be formed. In the main text, the angle $\angle(\t u,\t v)$ at the equator is used rather than $c$. These two are in a one-to-one correspondence
\[
    2 - 2 \cos(\t u,\t v) = g_{11} = \frac{(\cos\alpha+\cos\beta)^2}{c^2}.
\]
\section{Bending energy -- General case}
We have found that the admissible $\t x_{\mu\nu}$ satisfy five linear constraints: three describe the proportionality of $\t x_{11}$ and $\t x_{22}$, and two reduce to $\dd g_{12}=0$. Thus, the admissible $\t x_{\mu\nu}$ belong to a four-dimensional linear subspace of $(\R^3)^3$ space. Each dimension is spanned by a DOF that we attribute to one planarity defect $\delta^{(ij)}$, $(i,j)\in\{1,\bar 1\}^2$ such that
\[
\begin{split}
    \delta\t x^{(11)} &= \delta^{(11)}\t u\wedge\t s,\\
    \delta\t x^{(\bar11)} &= \delta^{(\bar11)}\t s\wedge\t v,\\
    \delta\t x^{(\bar1\bar1)} &= \delta^{(\bar1\bar1)}\t v\wedge\t t,\\
    \delta\t x^{(1\bar1)} &= \delta^{(1\bar1)}\t t\wedge\t u.
\end{split}
\]
Here too, we began by defaulting the same displacements as in~\eqref{eq:defZeroDisp} to zero. Next, we derive expressions for the $\delta^{(ij)}$ in function of the $\t x_{\mu\nu}$.

Generalizing equation~\eqref{eq:xmunuelem}, it is possible to derive full expressions for the $\t x_{\mu\nu}$ in terms of the $\delta^{(ij)}$. We find,
\[
\begin{split}
    \t x_{11}=
    &\frac{\scalar{\t u,\t s\wedge\t v}}{\scalar{\t t,\t s\wedge\t v}}
    \delta^{(11)}
    \t v\wedge\t t
    +
    \frac{\scalar{\t v,\t u\wedge\t s}}{\scalar{\t t,\t u\wedge\t s}}
    \delta^{(\bar11)}
    \t t\wedge\t u\\
    +
    &\frac{\scalar{\t v,\t t\wedge\t u}}{\scalar{\t s,\t t\wedge\t u}}
    \delta^{(\bar1\bar1)}
    \t u\wedge\t s
    +
    \frac{\scalar{\t u,\t v\wedge\t t}}{\scalar{\t s,\t v\wedge\t t}}
    \delta^{(1\bar1)}
    \t s\wedge\t v.
\end{split}
\]
As for $\t x_{12}$, it is readily available and reads
\[
\t x_{12} =
\delta^{(11)}\t u\wedge\t s
-
\delta^{(\bar11)}\t s\wedge\t v
+
\delta^{(\bar1\bar1)} \t v\wedge\t t
-\delta^{(1\bar1)}\t t\wedge\t u.
\]
These two vector equations provide a system (with two redundant equations), which can be solved for the $\delta^{(ij)}$. For instance, projecting $\t x_{11}$ over $\t u$ and $\t x_{12}$ over $\t v$ leads to a $2\times2$ system
\[
    \begin{split}
        \scalar{\t x_{11},\t u} &=
            \frac
            {\scalar{\t u,\t s\wedge\t v}
            \scalar{\t u,\t v\wedge\t t}}
            {\scalar{\t t,\t s\wedge\t v}}
            (\delta^{(11)}
            +
            \delta^{(1\bar1)}),\\
            \scalar{\t x_{12},\t v} &=
            \scalar{\t u,\t s\wedge\t v}\delta^{(11)}
            -
            \scalar{\t u,\t v\wedge\t t}\delta^{(1\bar1)},
    \end{split}
\]
which can be solved for $\delta^{(11)}$ and $\delta^{(1\bar1)}$ to give
\[\label{eq:deltas1}
\begin{split}
    \delta^{(11)}&=
    \frac{
    \scalar{\t x_{11},\t u}
    \scalar{\t t,\t s\wedge\t v}
    +
    \scalar{\t x_{12},\t v}
    \scalar{\t u,\t s\wedge\t v}
    }{
    \scalar{\t u,\t s\wedge\t v}
    \scalar{\t t-\t s,\t u\wedge\t v}
    },
    \\
    \delta^{(1\bar1)}&=\frac{
    \scalar{\t x_{11},\t u}
    \scalar{\t t,\t s\wedge\t v}
    -
    \scalar{\t x_{12},\t v}
    \scalar{\t u,\t v\wedge\t t}
    }{
    \scalar{\t u,\t v\wedge\t t}
    \scalar{\t t-\t s,\t u\wedge\t v}
    }.
\end{split}
\]
Similar considerations lead to
\[
\begin{split}
    \scalar{\t x_{11},\t v} &=
    \frac{\scalar{\t v,\t u\wedge\t s}\scalar{\t v,\t t\wedge\t u}}{\scalar{\t t,\t u\wedge\t s}}
    (\delta^{(\bar11)}+\delta^{(\bar1\bar1)}),\\
    \scalar{\t x_{12},\t u} &=
    -\scalar{\t u,\t s\wedge\t v}\delta^{(\bar11)}+
    \scalar{\t u,\t v\wedge\t t}\delta^{(\bar1\bar1)},
\end{split}
\]
and to
\[\label{eq:deltas2}
\begin{split}
    \delta^{(\bar11)}&=\frac{
    \scalar{\t x_{11},\t v}
    \scalar{\t t,\t u\wedge\t s}
    -
    \scalar{\t x_{12},\t u}
    \scalar{\t v,\t u\wedge\t s}
    }{
    \scalar{\t v,\t u\wedge\t s}
    \scalar{\t t-\t s,\t u\wedge\t v}
    },
    \\
    \delta^{(\bar1\bar1)}&=\frac{
    \scalar{\t x_{11},\t v}
    \scalar{\t t,\t u\wedge\t s}
    +
    \scalar{\t x_{12},\t u}
    \scalar{\t v,\t t\wedge\t u}
    }{
    \scalar{\t v,\t t\wedge\t u}
    \scalar{\t t-\t s,\t u\wedge\t v}
    }.
\end{split}
\]
Most importantly, the planarity defects $\delta^{(ij)}$ are linear forms of the full, in- and out-of-plane, components of the parametrization's second derivatives.

Now each panel in the unit cell, contributes a term $g^{(ij)}(r^2\delta^{(ij)})$, $(i,j)\in\{1,\bar 1\}^2$, to the bending energy density of the tessellation equal to the bending energy of the relevant panel for a given planarity defect $r^2\delta^{(ij)}$. The potential $g^{(ij)}$ can be nonlinear in principle, but given that the planarity defects are of order $O(r^2)$, it is reasonable to linearize it in the vicinity of $r\to 0$, that is while assuming, at the same time, that the plane is the natural state of the panels. In conclusion, the bending energy density of the tessellation takes the form
\[
b = \sum_{(i,j)\in\{1,\bar1\}^2}
\frac{1}{2}D^{(ij)}(\delta^{(ij)})^2
\]
where the $D^{(ij)}$ are the panels flexural rigidities normalized with respect to the area of a reference unit cell and where the $\delta^{(ij)}$ have been shown to be configuration-dependent (i.e., $(\t u,\t v,\t s,\t t)$-dependent) linear forms of the second derivatives $\t x_{\mu\nu}$. Equivalently, bending energy density $b=b(\Gamma_{i\mu\nu};g_{\alpha\beta})$ can be written as a metric-dependent quadratic form of the Christoffel symbols and of the curvatures. This is done next in the particular case of the ``Mars'' pattern.

Note that while we linearized $b$ into a quadratic form in the limit $r\to 0$ (i.e., for infinitesimal panel bending), we make no a priori assumptions regarding the macroscopic curvatures (i.e., the $\t x_{\mu\nu}$). These are free to take finite values. Last, the flexural rigidities $D^{(ij)}$ are material constants. Here, we suppose for simplicity that they are all equal to $D$.

\section{Bending energy -- The ``Mars'' pattern}
Similar considerations to the ones that led to equation~\eqref{eq:swtw} show that the current configuration of the folds $(\t u,\t v,\t s,\t t)$ can be determined from the tangent vectors $(\t x_1,\t x_2)$. Indeed, we have, in the case of an equilateral ``Mars'' pattern,
\[\begin{split}
    \t u &= \frac{1}{2}\t x_1 + \frac{\cos\alpha-\cos\beta}{g_{22}}\t x_2
    - u_n \t n,\\
    \t v &= -\frac{1}{2}\t x_1 + \frac{\cos\alpha-\cos\beta}{g_{22}}\t x_2
    - u_n \t n,\\
    \t s &= \frac{\cos\alpha+\cos\beta}{g_{11}}\t x_1+\frac{1}{2}\t x_2,\\
    \t t &= \frac{\cos\alpha+\cos\beta}{g_{11}}\t x_1-\frac{1}{2}\t x_2,
\end{split}
\]
with
\[
    u_n = \sqrt{1-\frac{g_{11}}{4}-\frac{(\cos\alpha-\cos\beta)^2}{g_{22}}}
\]
being a known function of the metric tensor. Thus, the coefficients appearing in the expression of $b$ can be written in terms of the metric tensor as well
\[
    \begin{split}
    \scalar{\t s,\t u\wedge\t v} &= -\scalar{\t t,\t u\wedge\t v} = u_n \sqrt{g}/2,\\
    \scalar{\t u,\t s\wedge\t t} &= \scalar{\t v,\t s\wedge\t t} =
    -\frac{\cos\alpha+\cos\beta}{g_{11}}\sqrt{g}u_n,
    \end{split}
\]
with $g=\det\t g=g_{11}g_{22}$. As for the remaining terms involving the $\t x_{\mu\nu}$, they can be expanded into combinations of the Christoffel symbols $\Gamma_{\sigma\mu\nu}\equiv\scalar{\t x_{\mu\nu},\t x_\sigma}$ and of the coefficients of the second fundamental form $\Gamma_{3\mu\nu}\equiv\scalar{\t x_{\mu\nu},\t n}$.
Namely, we have
\[
    \begin{split}
        \scalar{\t x_{11},\t u} &= \frac{1}{2}\Gamma_{111} + \frac{\cos\alpha-\cos\beta}{g_{22}}\Gamma_{211}
    - u_n \Gamma_{311},\\
    \scalar{\t x_{11},\t v} &= -\frac{1}{2}\Gamma_{111} + \frac{\cos\alpha-\cos\beta}{g_{22}}\Gamma_{211}
    - u_n \Gamma_{311},\\
    \scalar{\t x_{12},\t u} &= \frac{1}{2}\Gamma_{112} + \frac{\cos\alpha-\cos\beta}{g_{22}}\Gamma_{212}
    - u_n \Gamma_{312},\\
    \scalar{\t x_{12},\t v} &= -\frac{1}{2}\Gamma_{112} + \frac{\cos\alpha-\cos\beta}{g_{22}}\Gamma_{212}
    - u_n \Gamma_{312}.
    \end{split}
\]

Recall that, in the present case where the curvilinear coordinates are rectangular, the Christoffel symbols read
\[
    \begin{aligned}
    \Gamma_{111} &= \frac12g_{11,1},\quad &
    \Gamma_{211} &=-\frac12g_{11,2},\\
    \Gamma_{112}&= \frac12g_{11,2},\quad &
    \Gamma_{212}&=\frac12g_{22,1},\\
    \Gamma_{122}&=-\frac{1}{2}g_{22,1},\quad &
    \Gamma_{222}&=\frac{1}{2}g_{22,2}.
    \end{aligned}
\]

In conclusion, the total strain energy of a ``Mars'' pattern is
\[
    \psi(\t x) = \int_\Omega b(\gt\Gamma;\t g)\dd\xi_1\dd\xi_2
\]
where the bending strain energy density is
\[
  b(\gt\Gamma;\t g) =
  \frac{D}{2}\left((\delta^{(11)})^2+
  (\delta^{(\bar11)})^2+
  (\delta^{(1\bar1)})^2+
  (\delta^{(\bar1\bar1)})^2\right)
\]
with the planarity defects $\delta^{(ij)}$ being the metric-dependent linear forms given in~\eqref{eq:deltas1} and~\eqref{eq:deltas2}. Therein, note how the rigidity coefficients in front of the in-plane strain-gradient are proportional to the term $1/u_n$ which diverges when $g_{11}$ and $g_{22}$ approach their maximum and minimum values, i.e., at the boundary of the domain of kinematical admissibility. Meanwhile, the rigidity coefficients weighing the curvatures remain bounded.
\end{document}